\documentclass[sigconf,natbib=true]{acmart}

\copyrightyear{2023}
\acmYear{2023}
\setcopyright{acmcopyright}\acmConference[WSDM '23]{Proceedings of the Sixteenth ACM International Conference on Web Search and Data Mining}{February 27-March 3, 2023}{Singapore, Singapore}
\acmBooktitle{Proceedings of the Sixteenth ACM International Conference on Web Search and Data Mining (WSDM '23), February 27-March 3, 2023, Singapore, Singapore}
\acmPrice{15.00}
\acmDOI{10.1145/3539597.3570399}
\acmISBN{978-1-4503-9407-9/23/02}

\AtBeginDocument{%
  \providecommand\BibTeX{{%
    \normalfont B\kern-0.5em{\scshape i\kern-0.25em b}\kern-0.8em\TeX}}}

\usepackage{afterpage}
\usepackage{booktabs}
\usepackage{multirow}
\usepackage{algorithm}
\usepackage{algorithmic}
\usepackage{enumitem}
\usepackage{caption}
\usepackage{subcaption}
\usepackage{collab}
\usepackage{color}
\usepackage{adjustbox}
\usepackage{hyperref,tablefootnote,footnotehyper}

\setlist[itemize]{leftmargin=*}
\begin{document}

\title{A Bird's-eye View of Reranking: from List Level to Page Level}

\author{Yunjia Xi}
\authornote{Both authors contributed equally to this research.}
\email{xiyunjia@sjtu.edu.cn}
\affiliation{
  \institution{Shanghai Jiao Tong University}
  \city{Shanghai}
  \country{China}
}

\author{Jianghao Lin}
\authornotemark[1]
\email{chiangel@sjtu.edu.cn}
\affiliation{
  \institution{Shanghai Jiao Tong University}
  \city{Shanghai}
  \country{China}
}

\author{Weiwen Liu}
\authornote{The corresponding author.}
\email{liuweiwen8@huawei.com}
\affiliation{
  \institution{Huawei Noah's Ark Lab}
  \city{Shenzhen}
  \country{China}
}

\author{Xinyi Dai}
\email{daixinyi@sjtu.edu.cn}
\affiliation{
  \institution{Shanghai Jiao Tong University}
  \city{Shanghai}
  \country{China}
}

\author{Weinan Zhang}
\email{wnzhang@sjtu.edu.cn}
\affiliation{
  \institution{Shanghai Jiao Tong University}
  \city{Shanghai}
  \country{China}
}

\author{Rui Zhang}
\email{rayteam@yeah.net}
\affiliation{
  \institution{ruizhang.info}
  \city{Shenzhen}
  \country{China}
}

\author{Ruiming Tang}
\email{tangruiming@huawei.com}
\affiliation{
  \institution{Huawei Noah's Ark Lab}
  \city{Shenzhen}
  \country{China}
}

\author{Yong Yu}
\email{yyu@sjtu.edu.cn}
\affiliation{
  \institution{Shanghai Jiao Tong University}
  \city{Shanghai}
  \country{China}
}

\renewcommand{\shortauthors}{Yunjia Xi and Jianghao Lin, et al.}

\begin{abstract}
Reranking, as the final stage of multi-stage recommender systems, refines the initial lists to maximize the total utility. With the development of multimedia and user interface design, the recommendation page has evolved to a multi-list style. Separately employing traditional list-level reranking methods for different lists overlooks the inter-list interactions and the effect of different page formats, thus yielding suboptimal reranking performance. Moreover, simply applying a shared network for all the lists fails to capture the commonalities and distinctions in user behaviors on different lists. 
To this end, we propose to draw a bird's-eye view of \textbf{page-level reranking} and design a novel Page-level Attentional Reranking (PAR) model. 
We introduce a hierarchical dual-side attention module to extract personalized intra- and inter-list interactions. A spatial-scaled attention network is devised to integrate the spatial relationship into pairwise item influences, which explicitly models the page format. The multi-gated mixture-of-experts module is further applied to capture the commonalities and differences of user behaviors between different lists. Extensive experiments on a public dataset and a proprietary dataset show that PAR significantly outperforms existing baseline models.
\end{abstract} 



\maketitle

\section{Introduction}
\label{sec:intro}

In multi-stage recommender systems (MRS), reranking, as the final stage,  re-orders the input ranking lists from the previous ranking stage by modeling the cross-item influence~\cite{hron2021component}.
The goal of reranking is to maximize the total utility of the reranked lists.
The quality of reranking has a direct impact on users' experience and satisfaction, and thus plays a crucial role in MRS~\cite{liu2022neural}.

\begin{figure}[t]
    \centering
    \hfill
    \begin{subfigure}{0.11\textwidth}
        \centering
        \includegraphics[width=0.98\linewidth]{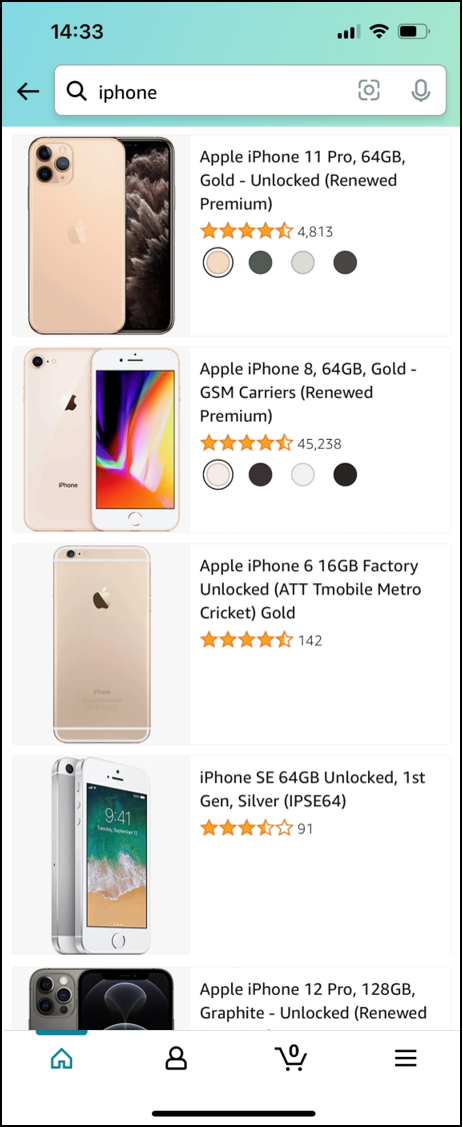}
        \caption{}
        \label{fig:page format illustration a}
    \end{subfigure}
    \hfill
    \begin{subfigure}{0.11\textwidth}
        \centering
        \includegraphics[width=0.98\linewidth]{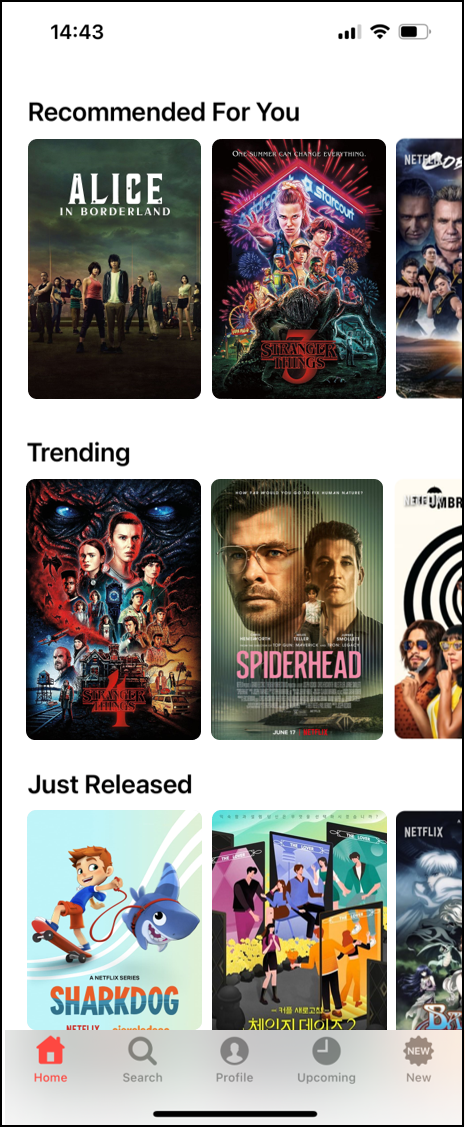}
        \caption{}
        \label{fig:page format illustration b}
    \end{subfigure}
    \hfill
    \begin{subfigure}{0.11\textwidth}
        \centering
        \includegraphics[width=0.98\linewidth]{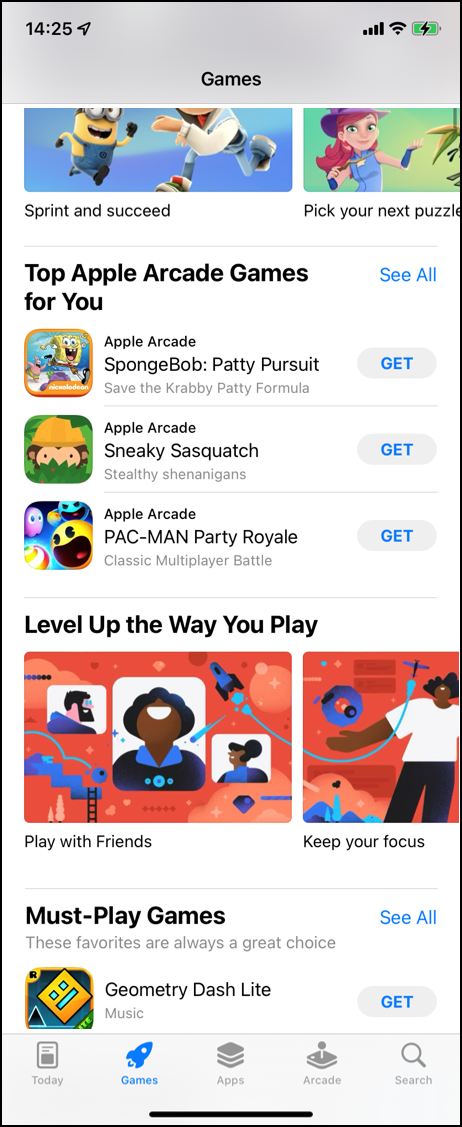}
        \caption{}
        \label{fig:page format illustration c}
    \end{subfigure}
    \hfill
    \begin{subfigure}{0\textwidth}
        \centering
        \includegraphics[width=0.98\linewidth]{figures/page-format-no-arrows-c.png}
    \end{subfigure}
    \vspace{-10pt}
    \caption{Illustration of different page formats. Left: single vertical list in Amazon Shopping. Medium: multiple horizontal lists in Netflix. Right: multi-list page with interleavings of vertical and horizontal lists in Apple App Store.}
    \label{fig:page format illustration}
    \vspace{-15pt}
\end{figure}


Various reranking methods~\cite{ai2018learning,pang2020setrank,pei2019personalized,ai2019learning,zhuang2018globally} have been developed in recent years, but they are mainly \textit{list-level reranking} models. List-level models rerank a single list each time, and only consider the cross-item influence within the individual list~\cite{liu2022neural}. Such a reranking strategy, though already found useful in many industrial applications, may still be suboptimal. In fact, with the development of multimedia and user interface design, the final recommendation page presented to the user is usually in a \textit{multi-list style}~\cite{gomez2015netflix,fu2022f}. As shown in Figure~\ref{fig:page format illustration}, each list on the multi-list page often highlights a particular theme (e.g., ``Trending'', ``Games''), sometimes even with a tailored layout (e.g., size, location).
The existence of other lists changes the user behavior patterns, leading to a different utility distribution. 
In this work, we propose to draw a bird's-eye view over the whole reranking page and develop a \textit{page-level reranking} algorithm. While intuitively useful to integrate page-wise information, it is nontrivial to jointly perform reranking for multiple lists with the following three major challenges (\textbf{C1} - \textbf{C3}).

\textbf{(C1)} Firstly, incorporating intra-list and inter-list interactions is essential for page-level reranking. Items from the same list are usually related to the same theme. Modeling the cross-item influence within the list (i.e., intra-list interaction) and identifying the best permutation over candidate items is the main objective for reranking~\cite{liu2022neural}. 
Moreover, we observe that the inter-list interaction is also crucial for page-level modeling --- whether a user is interested in an item is also influenced by items placed in other lists. Independent optimization for each individual list ignores the context in other lists. 
For example, in Figure~\ref{fig:page format illustration b}, the TV series \emph{Stranger Things Season 3} should have a low utility in the ``Recommended For You'' list (rank 2), supposing the user had already finished the series years ago.
However, its sequel \emph{Stranger Things Season 4} in the ``Trending'' list (rank 1) can motivate the user to re-watch the \emph{Season 3} for a second time to recall the previous story, leading to a different utility distribution.
Thus, both intra- and inter-list interactions should be considered to provide a holistic view.


\textbf{(C2)} Secondly, the page format of the reranking page affects how items interact with each other, and thus should also be introduced to the page-level reranking. For instance, in Figure~\ref{fig:page format illustration b}, lists of movies and TV shows are horizontally stacked from top to bottom. In Figure~\ref{fig:page format illustration c}, the page contains interleavings of vertical and horizontal lists and forms an ``F'' shape. Compared with the page of stacked horizontal lists, the horizontal lists in the F-shape page are separated by a vertical list with a larger distance, so that the influence between items from two consecutive horizontal lists may be less. The influence becomes even less if the length or size of the inserted vertical list is increased, which further enlarges the distance between lists. We also provide evidence for such an effect in Section~\ref{sect:data_analysis}. Therefore, page-level reranking is expected to formulate the page format (e.g., the size and location of the items).


 \begin{figure*}[h]
    \centering
    \begin{minipage}[c]{0.12\textwidth}
    \includegraphics[width=\linewidth]{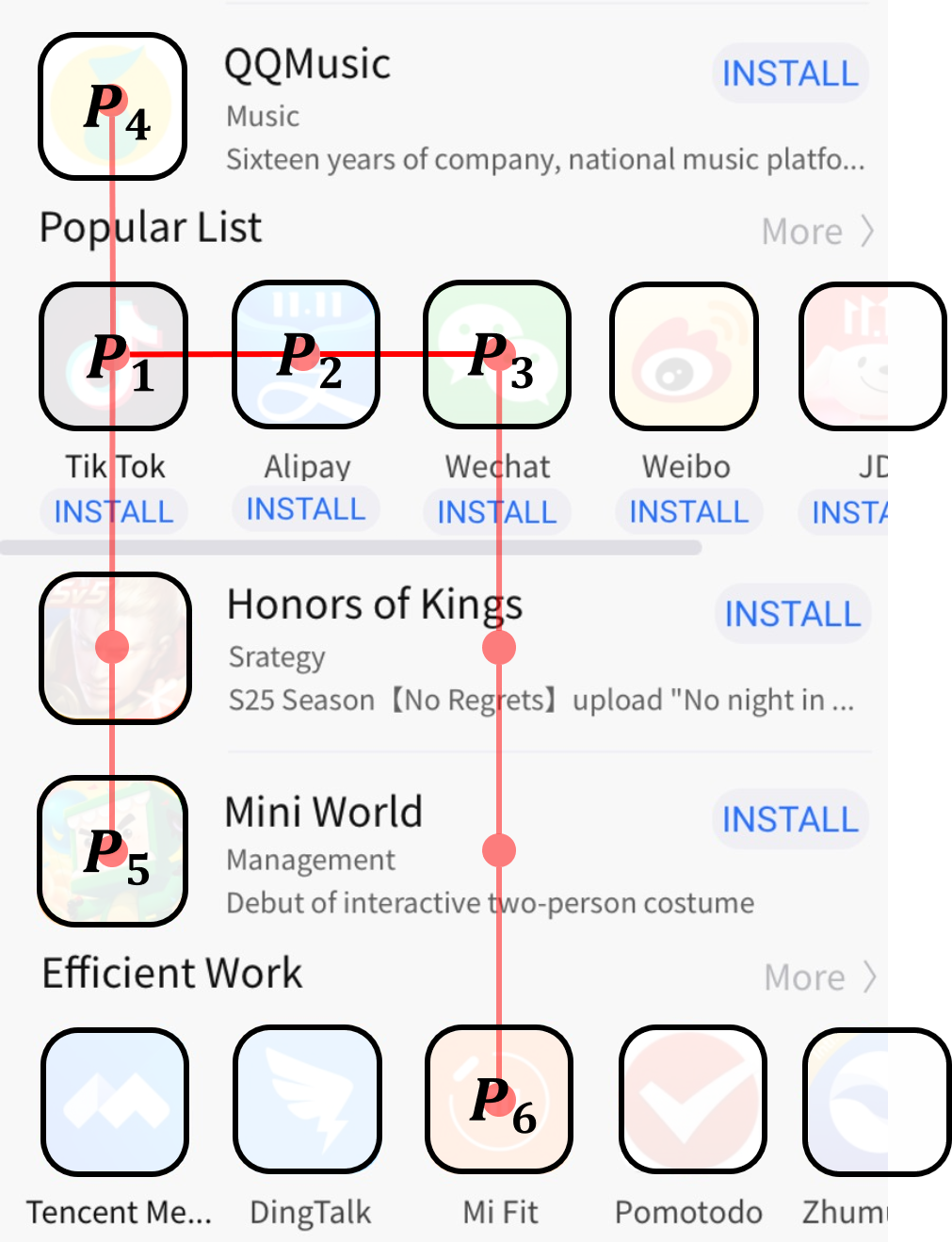}
    \subcaption{}
    \label{fig:data_analysis_illustr}
    \end{minipage}
    \hspace{7pt}
    \begin{minipage}[c]{0.16\textwidth}
    \includegraphics[width=\linewidth]{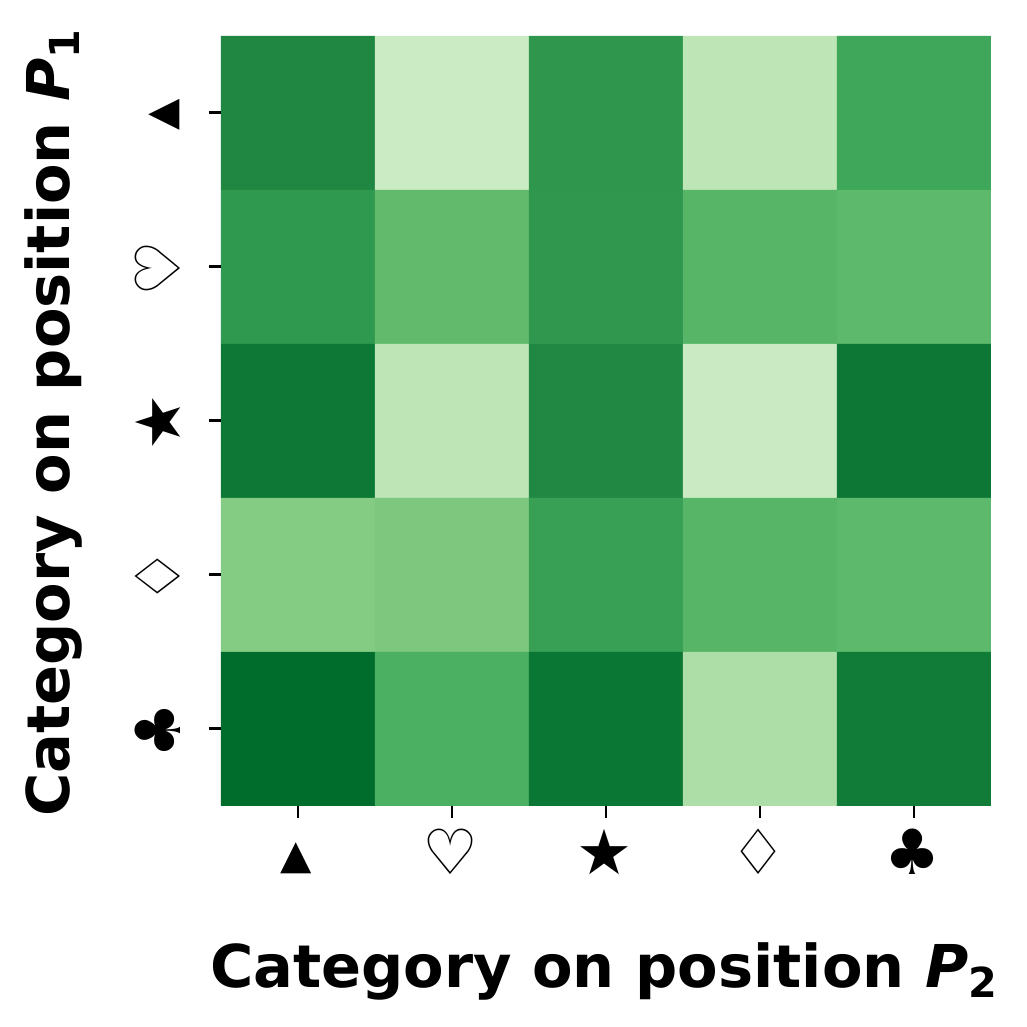}
    \subcaption{}
    \label{fig:pos12}
    \end{minipage}
    \begin{minipage}[c]{0.16\textwidth}
    \includegraphics[width=\linewidth]{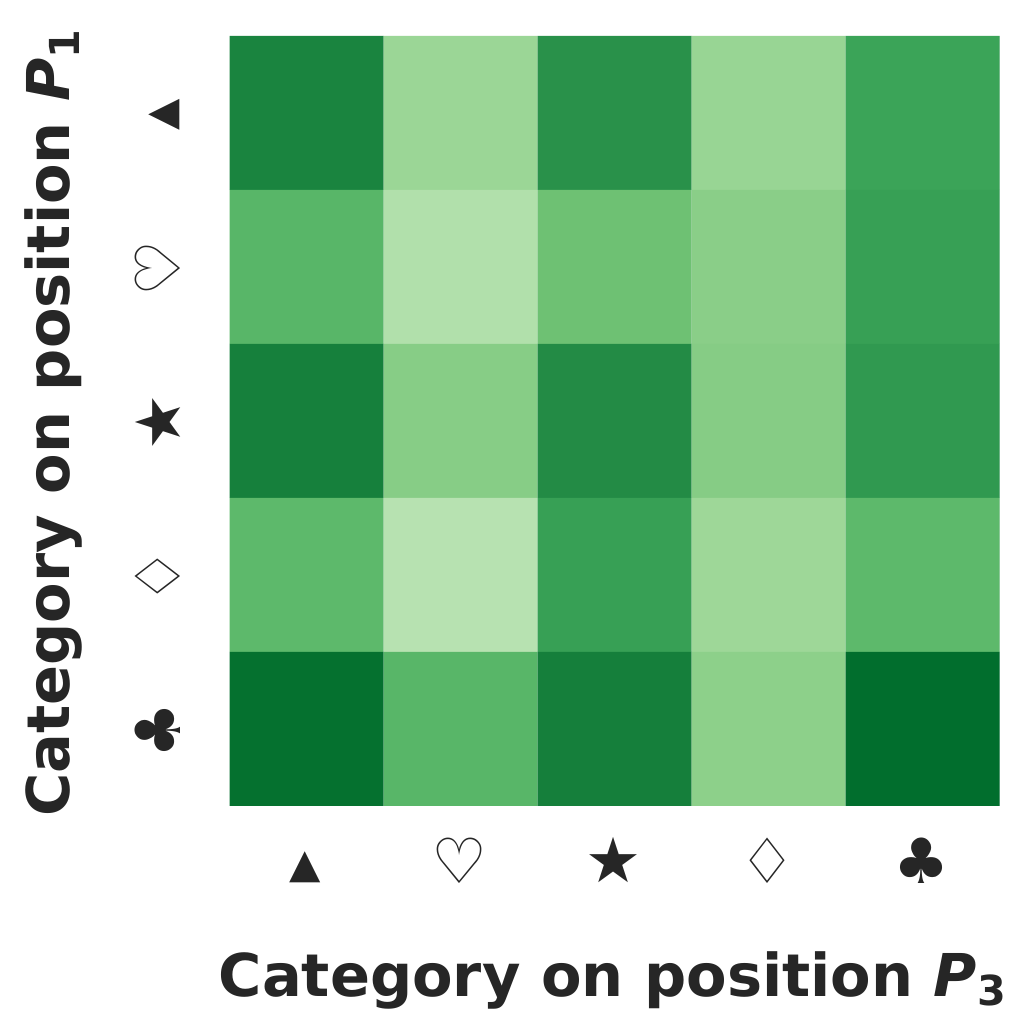}
    \subcaption{}
    \label{fig:pos13}
    \end{minipage}
    \begin{minipage}[c]{0.16\textwidth}
    \includegraphics[width=\linewidth]{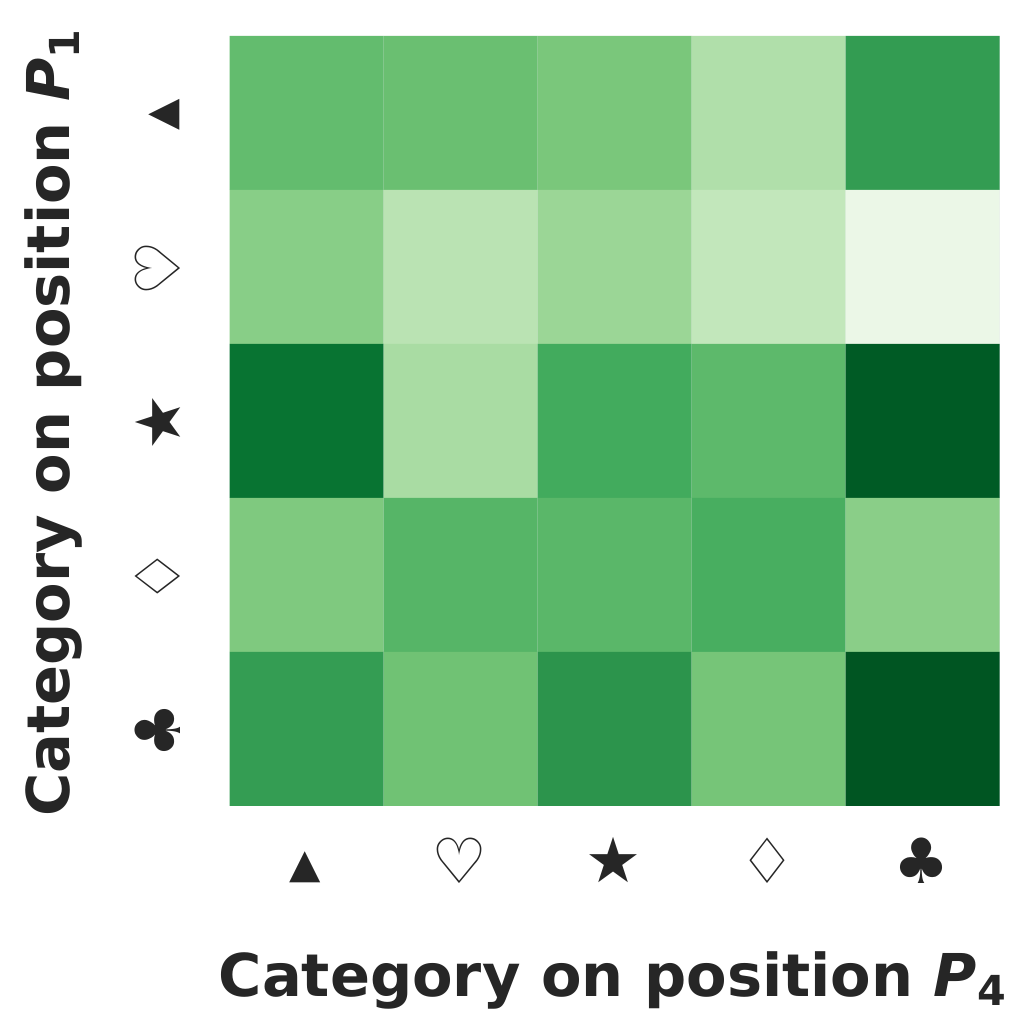}
    \subcaption{}
    \label{fig:pos14}
    \end{minipage}
    \begin{minipage}[c]{0.16\textwidth}
    \includegraphics[width=\linewidth]{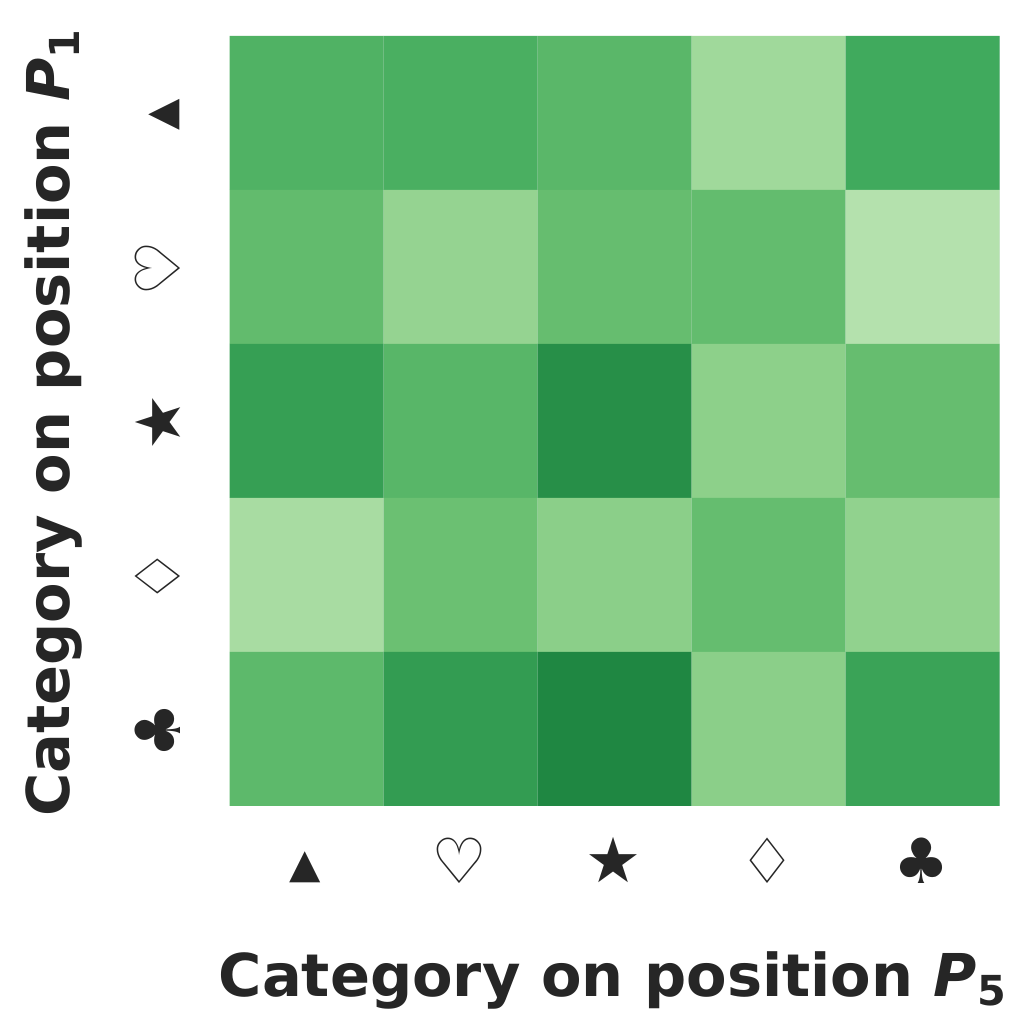}
    \subcaption{}
    \label{fig:pos15}
    \end{minipage}
    \begin{minipage}[c]{0.195\textwidth}
    \includegraphics[width=\linewidth]{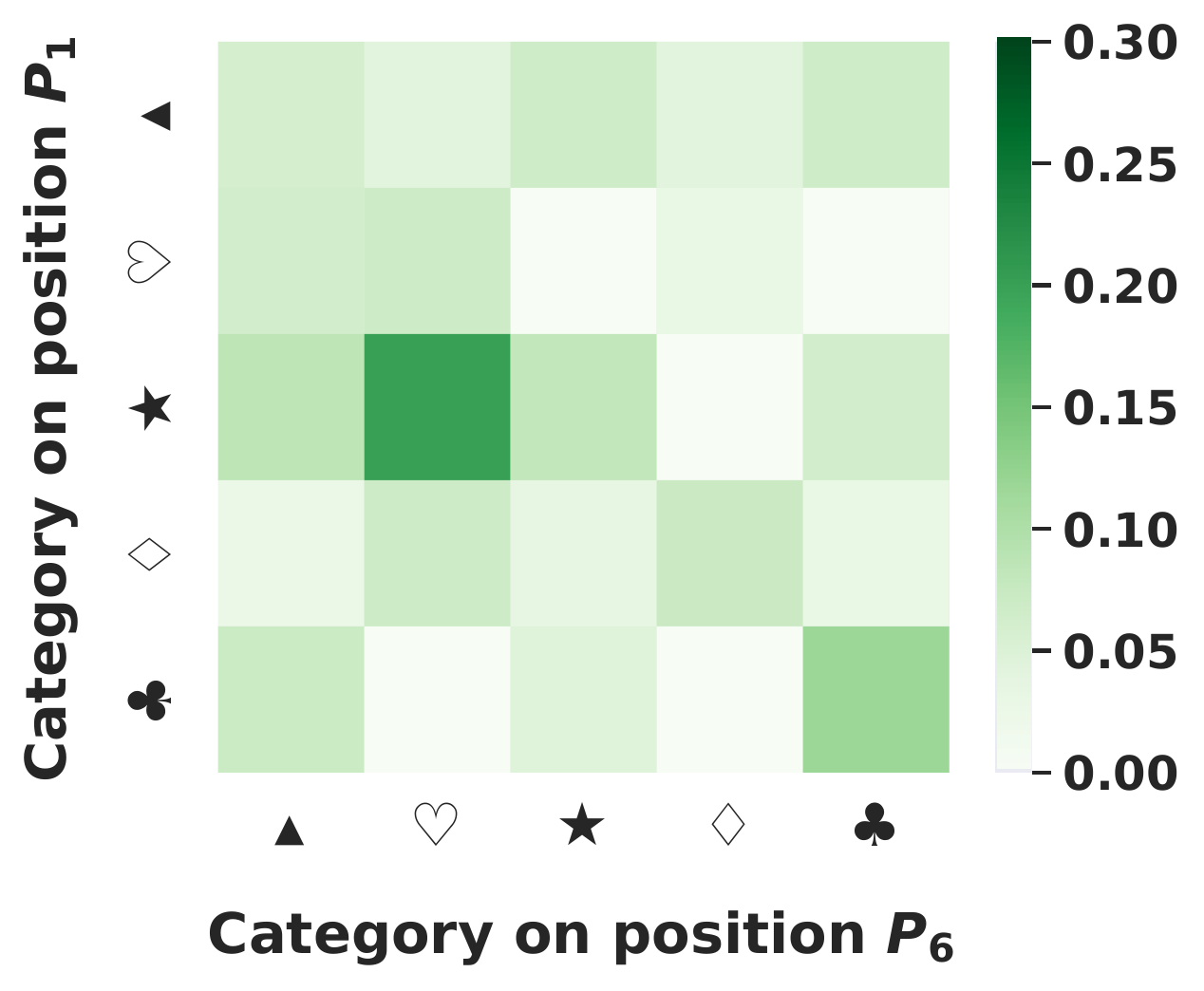}
    \subcaption{}
    \label{fig:pos16}
    \end{minipage}
    \vspace{-10pt}
    \caption{(a) The five fixed positions on the F-shape page. The red auxiliary lines and dots are provided for the illustration of the Manhattan distance measurement between items. 
    (b) CTR of $P_1$ as $P_2$ varies. (c) CTR of $P_1$ as $P_3$ varies. (d) CTR of $P_1$ as $P_4$ varies. (e) CTR of $P_1$ as $P_5$ varies. (f) CTR of $P_1$ as $P_6$ varies.}
    \label{fig:data_analysis}
    \vspace{-5pt}
\end{figure*}

\textbf{(C3)} Thirdly, user behaviors on different lists have commonalities and distinctions. For one thing, user behaviors across lists may share some basic patterns (e.g., position bias, cascade hypothesis~\cite{craswell2008experimental,Guo2009CCM}) and underlying personal preferences. Collaboratively putting multiple lists together benefits the understanding of the underlying common behaviors of each user. For another, due to the theme and format of each list, user behaviors also possess distinctions for different lists. For example, in Figure~\ref{fig:page format illustration c}, the position bias could be more severe for horizontal lists than that for vertical ones. The horizontal lists are designed as carousel sliders, which reduces the impression opportunity for lower-ranked items. Simply using the same network for all the lists may be inferior as the list-specific information is not well captured, especially for low-resource or sparse reranking lists. 

Although \citet{hao2021re} consider the page-level information and propose the DHANR model, they only transform the items on the page to a static page representation by a hierarchical attention network. The obtained page representation is fed into the list-level reranking model as the static side information for each individual list, unaware of the dynamic inter-list interactions between items or the page format of the whole page. Moreover, DHANR fails to capture the commonalities and distinctions among lists.

To address the above issues, we propose a novel model named \underline{P}age-level \underline{A}ttentional \underline{R}eranking (PAR) for page-level reranking. Multiple lists are jointly reranked with a unified model to capture the multifaceted fine-grained mutual influences among lists. Firstly, we propose hierarchical dual-side attention (HDS-Attn) module to extract the intra- and inter-list interactions according to users' individual behavior history \textbf{(C1)}. 
Next, the spatial-scaled attention (SS-Attn) network is designed to encapsulate the pairwise influence between items with respect to their spatial relationship. The attention is numerically scaled by the distance between items on the page, which provides explicit modelings over the page format \textbf{(C2)}. 
Lastly, after obtaining the
interacted feature representation from HDS-Attn and SS-Attn, PAR adopts the Multi-gated Mixture-of-Experts (MMoE~\cite{ma2018modeling}) module to capture the commonalities and differences of user behavior patterns among different lists \textbf{(C3)}. MMoE maintains a set of parallel expert networks to capture various aspects of behavior patterns and applies an attentional gate with list-specific parameters for each individual list to aggregate the expert outputs for the final score estimation.

The main contributions of this paper are listed as follows:
\begin{itemize}
    \item We propose to draw a bird’s-eye view over the whole reranking page to jointly rerank multiple lists and optimize the overall utility by considering the page-wise information. To the best of our knowledge, this is the first work to consider the effect of general page format for page-level reranking.
    \item We conduct data analysis based on a multi-list dataset and identify the importance of intra- and inter-list interactions and the spatial relationship (i.e., reflection of page format) between items.
    \item We propose a novel Page-level Attentional Reranking (PAR) model. We design an HDS-Attn module for the personalized intra- and inter-list interactions, and an SS-Attn module to incorporate the page format. The MMOE module is equipped to capture the commonalities and distinctions for different lists.
    \item Extensive experiments on a public dataset and a proprietary dataset show that PAR achieves the state-of-the-art performance compared with existing baseline models.
\end{itemize}

\section{Data Analysis}
\label{sect:data_analysis}


One of the challenges for page-level reranking is to model the intra- and inter-list interactions between items on a page. Therefore, we present a brief data-driven study on an F-shape page (i.e., Figure~\ref{fig:data_analysis_illustr}) to investigate how items at different positions influence each other. The data is collected from a mainstream App Store, where the page adopts an interleaved arrangement of vertical and horizontal lists. The detailed data description can be found in Section~\ref{sec:dataset}.

To study how the other items placed on the same page influence the utility of a given item, we select 6 fixed positions on the F-shape page, as illustrated in Figure~\ref{fig:data_analysis_illustr}. For the fixed position $P_1$:
\begin{itemize}[leftmargin=10pt]
    \item $P_2$ is adjacent to $P_1$ in the same list;
    \item $P_3$ is distant from $P_1$ in the same list;
    \item $P_4$ is adjacent to $P_1$ in another neighboring list;
    \item $P_5$ is distant from $P_1$ in another neighboring list;
    \item $P_6$ is distant from $P_1$ in a remote list.
\end{itemize}

They form five typical positional relationships on a multi-list page. For each pair of positions (i.e., $P_1$ versus $P_i$, $i=2,3,4,5,6$), we compute the click-through rate (CTR) of $P_1$ with different categories when the item category at $P_i$ varies. We plot the heatmap for the CTR w.r.t. the top five most frequent categories in Figure~\ref{fig:pos12} to \ref{fig:pos16}. Each symbol at the horizontal and vertical axes denotes an item category. The color variation represents how the CTR changes with the varying of the item category, and thereby reflects the items' mutual influences. From the figures, we have the following central observations (\textbf{Obs. I} to \textbf{Obs. III}):


\textbf{Obs. I:} \textit{Item utility is influenced by other items in the same list.} The greater the variation of the color in the heatmap, the stronger the influence between the items.  Figure~\ref{fig:pos12} ($P_2$) and~\ref{fig:pos13} ($P_3$) display how the CTR changes with the variation of the other items in the same list. 
It is evident that if the category of $P_1$ is fixed, the CTR of $P_1$ will vary with the category of $P_2$ and $P_3$. 


\textbf{Obs. II:} \textit{Item utility is influenced by other items across different lists.} Figure~\ref{fig:pos14} to \ref{fig:pos16} show the inter-list interactions between items. 
The CTR of $P_1$ in Figure~\ref{fig:pos14} to \ref{fig:pos16} varies with different categories in the other position, verifying the item influence across lists. 
The color variation in the heatmap reflects the change in CTR, and further indicates the impacts of items. 
In particular, although $P_4$ (Figure~\ref{fig:pos14}) is placed in the different lists, its impact on $P_1$ is similar to $P_2$ (Figure~\ref{fig:pos12}) from the same list, illustrating that inter-list interactions can be comparable to intra-list interactions for multi-list pages.

\textbf{Obs. III:} \textit{The influence between items shows a negative correlation with the distance between items.} We adopt the Manhattan distance to measure the distance between two items to study how the distance affects the influence between them. The Manhattan distance is the sum of the difference on horizontal and vertical axes, as illustrated by the red auxiliary \emph{lines} and \emph{dots} in Figure~\ref{fig:data_analysis_illustr}. The distances from $P_1$ to $P_i\,(i=2,3,4,5,6)$ are $1$, $2$, $1$, $2$, $5$, respectively. Compared with Figure~\ref{fig:pos12}~($P_2$), the variation of the color in Figure~\ref{fig:pos13} ($P_3$) is more uniform, which suggests that the impact of a distant item is less than the adjacent ones in the same list. Similarly, for different lists, the CTR in Figure~\ref{fig:pos15} ($P_5$) and Figure~\ref{fig:pos16} ($P_6$) also varies more evenly than that in Figure~\ref{fig:pos14} ($P_4$). The light color of Figure~\ref{fig:pos16} is probably due to the fact that the dataset only records the observed data by the user, and the lower-ranked position $P_6$ is less observed with few data records when $P_1$ is clicked.

Therefore, we are motivated to propose a page-level reranking model that is aware of intra- and inter-list interactions and the spatial relationships between items across lists. 

\section{Methodology}

\begin{figure*}[t]
    \centering
    \includegraphics[width=0.99\textwidth]{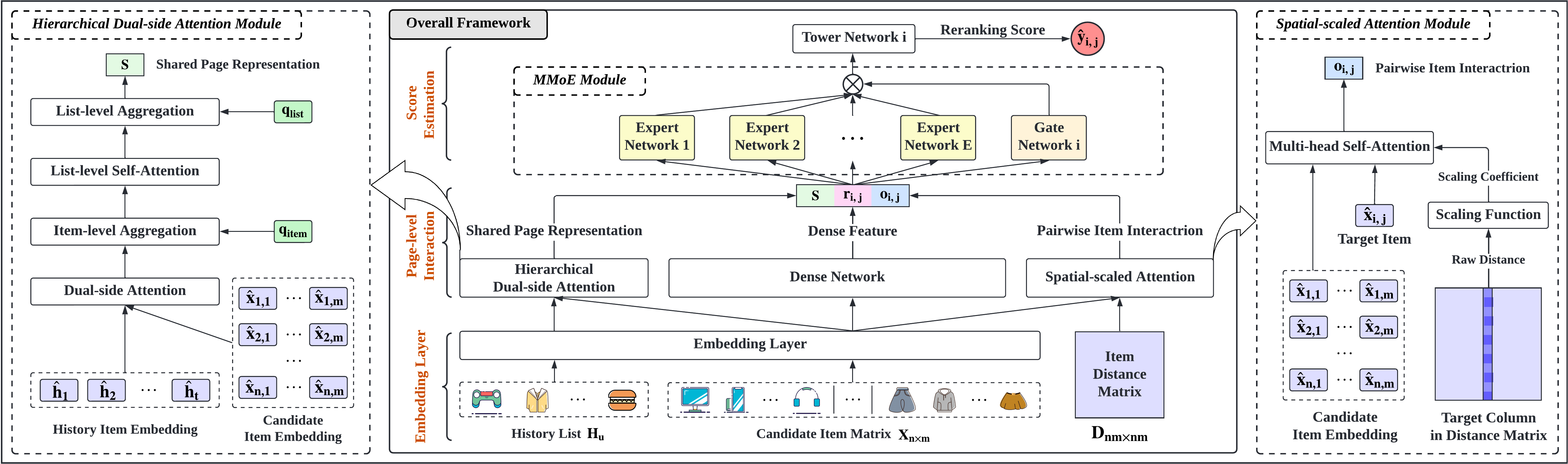}
    \vspace{-5pt}
    \caption{The overall framework of PAR, which consists of three major layers: (1) embedding layer, (2) page-level interaction layer, and (3) score estimation layer.}
    \vspace{-5pt}
    \label{fig:framework}
\end{figure*}
\subsection{Problem Formulation}

A page-level reranking model takes as inputs the multiple ordered initial lists on the same page generated by the previous rankers, and refines the ranking lists by considering the mutual influence between items and the impact of the page format.

For a user $u$ with the associated history list $H_u=[h_1^u,\ldots,h_t^u]$, and a multi-list page $P\in\mathcal{P}$ of $n$ different initial lists $\{R_1,R_2,\dots,R_n\}$, the problem of the page-level reranking is to jointly rerank the $n$ lists to optimize the overall utility by considering the page-wise context. 
Here, $t$ is the maximum length of the user's history list, $h_s^u$, $s=1\ldots,t$ is the $s$-th item recently clicked by the user $u$, $\mathcal P$ is the page set, and $R_i$, $i=1,\ldots,n$ is the $i$-th list with $m_i$ items. 
Let $m$ be the maximum length of the lists on the page. Then the whole page of candidate items to be reranked can be represented by an item matrix $\mathbf{X}_{n\times m}$, where $x_{i,j}$ denotes the $j$-th item in the $i$-th list, with all lists padded to the maximum length $m$. The utility of a page is defined by the expected sum (or weighted sum) of the click\footnote{The click could be replaced by other utility metrics like conversions, purchases, etc.} probability for each item on the page.


\label{sec:framework}

\vspace{-5pt}
\subsection{Model Overview}
We present Page-level Attentional Reranking (PAR) model and introduce how PAR captures the multifaceted fine-grained interactions for the whole page. As depicted in Figure~\ref{fig:framework}, PAR is an end-to-end reranking model, consisting of three layers: (1) embedding layer, (2) page-level interaction layer, and (3) score estimation layer.

\emph{Firstly}, the embedding layer converts each candidate or historical item to a dense feature vector. 
\emph{Then}, in the page-level interaction layer, we design three modules (i.e., hierarchical dual-side attention, spatial-scaled attention, and dense network) to capture page-wise information. 
\emph{Finally}, we adopt Multi-gated Mixture-of-Experts to learn the commonalities and distinctions among different lists and output the predicted score in the score estimation layer.
\vspace{-8pt}
\subsection{Embedding Layer}

PAR applies the embedding layer to transform the sparse raw features into low-dimensional dense embedding vectors. The embedding layer takes as input both the candidate item matrix and the user's history list. Specifically, let $\hat{x}_{i,j}\in \mathbb R^{d_x}$ be the embedding of the item $x_{i,j}$, $i=1,\ldots,n$, $j=1,\ldots,m$ from the candidate item matrix, and $\hat h_s\in \mathbb R^{d_h}$ be the embedding of the item $h_s$, $s=1,\ldots,t$ from the history list, where $d_x$ and $d_h$ are the embedding size for the candidate and history items, respectively. 
Then, we obtain the candidate item embedding matrix $\hat{\mathbf{X}}\in\mathbb{R}^{n\times m\times d_x}$ and the history item embedding matrix $\hat{\mathbf{H}}\in\mathbb{R}^{t\times d_h}$ after the embedding layer, where $t$ is the history length, $n$ is the number of lists, and $m$ is the number of items in each list.

\subsection{Page-level Interaction}

To fully exploit the page-wise information, we propose three modules in the page-level interaction layer: (i) The hierarchical dual-side attention (HDS-Attn) module learns the personalized intra- and inter-list interactions, and generates a shared page representation; (ii) The spatial-scaled attention (SS-Attn) module models the pairwise fine-grained item influence with the awareness of the page format effect; (iii) The dense network captures implicit feature interactions within each item. Detailed descriptions for each module are given in the following sections.

\subsubsection{Hierarchical Dual-side Attention}

One of the challenges of page-level reranking is to incorporate intra- and inter-list interactions on a page (\textbf{C1} in Section~\ref{sec:intro}). Since there is a natural hierarchical structure of \emph{items forming a list} and \emph{lists forming a page}, we propose to consider intra-list and inter-list interaction in a hierarchical way. As shown in the left part of Figure~\ref{fig:framework}, the HDS-Attn module consists of dual-side attention, item-level aggregation, list-level self-attention, and list-level aggregation, from bottom to top. The bottom two parts are designed to extract the intra-list interaction within each list. To provide personalized reranking, the cross-item influence on the candidate item side requires the personal preference information from the history list side. Therefore, we introduce the dual-side attention for each list to model the mutual influence between these two sides. Then, item-level aggregation is employed to combine the item information within a list and generate \textit{list representation}. The upper two parts, list-level self-attention and list-level aggregation, are designed to capture the inter-list interaction between lists and generate the final \textit{page representation}. 


\smallskip
\noindent \textbf{Dual-side Attention.} Users' history list carries rich information for inferring their personal preferences and tastes, which is helpful for reranking~\cite{xi2022multi,li2022pear}. Moreover, the items in the users' history list contribute differently for different candidate lists~\cite{xi2022multi,qin2020user}. For example, game apps in the user's history list may be more critical when reranking the "Top-10 Games" list for the user. Inspired by~\cite{co_att}, we design the dual-side attention to model the fine-grained correlations between the candidate item side and history list side.

Formally, for each list $R_i$ on page $P$, $i=1,\ldots,n$, the input of the dual-side attention is the corresponding $i$-th item embedding matrix $\hat{\mathbf{X}}_i\in\mathbb{R}^{m\times d_x}$ of the candidate list $R_i$ and the history embedding matrix $\hat{\mathbf{H}}\in\mathbb{R}^{t\times d_h}$.
We maintain an affinity matrix $\mathbf{W}_i^a\in\mathbb{R}^{d_h \times d_x}$ for each list to learn the importance of the association between each pair of items from both candidate and history sides:
\begin{equation}
\begin{split}
    \mathbf{C}_i&=\tanh{(\hat{\mathbf{H}}\mathbf{W}_i^{a}\hat{\mathbf{X}}_i^\top )}\,, \\
    \mathbf{A}_i^x&=\operatorname{Softmax}\left(\tanh{(\hat{\mathbf{X}}_i\mathbf{W}_i^x+(\hat{\mathbf{H}}\mathbf{W}_i^h)^\top \mathbf{C}_i)}\right),\\
    \mathbf{A}_i^h&=\operatorname{Softmax}\left(\left(\tanh{(\hat{\mathbf{H}}\mathbf{W}_i^h + \mathbf{C}_i(\hat{\mathbf{X}}_i\mathbf{W}_i^x))}\right)^\top\right),\\
    \Tilde{\mathbf{X}}_i&=\mathbf{A}_i^x\hat{\mathbf{X}}_i\,,\;\;\;\;\;\Tilde{\mathbf{H}}_i=\mathbf{A}_i^h \hat{\mathbf{H}}\,,
\end{split}
\label{eq:dual-side attention}
\end{equation}
where $\mathbf{W}_i^x\in\mathbb{R}^{d_x\times m}, \mathbf{W}_i^h\in\mathbb{R}^{d_h\times m}$ are learnable weight matrices. 
The matrices $\mathbf{A}_i^x\in\mathbb{R}^{m\times m}$ and $\mathbf{A}_i^h\in\mathbb{R}^{m \times t}$ after the softmax function represent the attention weights of items in the candidate list and history list. 
Then the interacted representation matrices $\Tilde{\mathbf{X}}_i=[\Tilde{\mathbf{x}}_{i,j}]_{j=1}^m\in\mathbb{R}^{m\times d_x}$ and $\Tilde{\mathbf{H}}_i=[\Tilde{\mathbf{h}}_{i,j}]_{j=1}^m\in\mathbb{R}^{m\times d_h}$ now contain useful information from both candidate list $R_i$ and history list $H_u$. 

\smallskip
\noindent \textbf{Item-level Aggregation.}
We apply the item-level aggregation to learn the intra-list interaction and generate the list representations.
Since items contribute differently to the representation of the target list $R_i$, we aggregate the attained item representations with an attention mechanism \cite{yang-etal-2016-hierarchical} to form the \textit{list representation} $l_i$:
\begin{equation}
\begin{split}
    u_{i,j} &= \tanh{(\mathbf{W}_l[\Tilde{\mathbf{x}}_{i,j} \,\Vert\, \Tilde{\mathbf{h}}_{i,j}] + b_l)}\,,\\
    \alpha_{i,j}& =\frac{\text{exp}(u_{i,j}^\top  q_{\text{item}})}{\sum_{j'=1}^{m} \text{exp}(u_{i,j'}^\top q_{\text{item}})}\,,\\
    l_i& =\sum\nolimits_{j=1}^m{\alpha_{i,j}[\Tilde{\mathbf{x}}_{i,j} \,\Vert\, \Tilde{\mathbf{h}}_{i,j}]}\,,\\
\end{split}
\end{equation}
where $\Vert$ denotes the vector concatenation.
We first feed the concatenated item representations into a linear layer to get $u_{i,j}$ for each item $x_{i,j}$, where $\mathbf{W}_l\in\mathbb R^{d_l\times d_l}$ and $b_l\in\mathbb R^{d_l}$ ($d_l=d_x+d_h$) are the learnable weights. Then the importance of each item is measured by the similarity of $u_{i,j}$ with a item-level query vector $q_{\text{item}}\in\mathbb{R}^{d_l}$. The item-level query vector $q_{\text{item}}$ is a trainable parameter and serves as the attention query in the item-level aggregation. Next, we normalize the weights $\alpha_{i,j}$ and compute the list representation $l_i$ by the weighted sum of each item. To this end, important intra-list interaction has been fused into the list representation $l_i$. Stacking all the list representations, we get the list representation matrix $\textbf{L}=[l_i]_{i=1}^n\in\mathbb{R}^{n\times d_l}$ for further use. 

\smallskip
\noindent \textbf{List-level Self-attention.}
Given the list representation matrix $\textbf{L}$, we model the inter-list influence between different lists on the page through a list-level self-attention layer:
\begin{equation}
    \Tilde{\mathbf{L}}=\text{Softmax}\left(\frac{\mathbf{L}\mathbf{L}^\top}{\sqrt{d_l}}\right)\mathbf{L}\,,
\end{equation}
where $\Tilde{\mathbf{L}}=[\Tilde{l}_i]_{i=1}^n \in \mathbb R^{n\times d_l}$ is the re-weighted list representation matrix that captures the relationship and correlations between different lists, and $\sqrt{d_l}$ is used to stabilize gradients during training.

\smallskip
\noindent \textbf{List-level Aggregation.}
Finally, built upon the re-weighted list representations $\Tilde{\mathbf{L}}$, we employ the list-level aggregation layer to combine the information from different lists and generate the unified page representation $S$. Similar to the item-level aggregation, after a linear transformation,  we involve a learnable list-level query vector $q_{\text{list}}\in\mathbb{R}^{d_l}$ for calculating the attention weights $\beta$. Then the re-weighted list representations $\Tilde{\textbf{L}}$ are aggregated into a shared \textit{page representation} $S$ by a weighted sum:
\begin{equation}
\begin{split}
    v_i& = \tanh{(W_p \Tilde{l}_i + b_p)}\,,\\
    \beta_i& =\frac{\text{exp}(v_i^\top q_{\text{list}})}{\sum_{i'=1}^n \text{exp}(v_{i'}^\top q_{\text{list}})}\,,\\
    S& =\sum\nolimits_{i=1}^n{\beta_i \Tilde{l}_i}\,,\\
\end{split}
\end{equation}
where page representation $S$ integrates the information of page-wise contexts and history behaviors, and is shared for all the lists.

\subsubsection{Spatial-scaled Attention}
\label{sec:spatial-scaled attention}
With multiple lists on one page, the arrangement of lists and items, i.e., the page format, becomes an essential issue to be considered (\textbf{C2} in Section~\ref{sec:intro}). 
However, no previous work on page-level reranking has discussed such an issue before. Different page formats change the location of items and therefore change the distances between them. In Section~\ref{sect:data_analysis}, we observe that the influence between items generally shows a negative correlation with the distance between them. Hence different page formats yield diverse influences between items. 
Thus, we propose the SS-Attn module to estimate the pairwise item influence with the consideration of the spatial relationship. Specifically, SS-Attn adjusts the attention weights according to the relative distance between items on the page, so that closer items have a stronger influence on each other. The relative distance can be altered depending on the page format, showing that the SS-Attn is flexible and can be adapted to different formats.


As such, with a total of $nm$ items on the page, we introduce a symmetric distance matrix $\mathbf{D}\in\mathbb{R}^{nm \times nm}$, whose element $d_{p, q}\ge 0$ indicates the geometric distance between the corresponding pair of items $(p, q)$. In this work, we adopt the Manhattan distance to build the distance matrix. 
For example, in Figure~\ref{fig:data_analysis_illustr}, the Manhattan distances from $P_1$ to $P_i\,(i=2,3,4,5,6)$ are $1$, $2$, $1$, $2$, $5$, respectively.
It is worth noting that the design of the distance matrix is flexible and can be customized according to the page formats in different page-level reranking scenarios. Other distance measurements (e.g., Euclidean distance, slot counting) are also applicable.

Concretely, we first reshape the embedding matrix of the candidate items $\hat{\mathbf{X}}\in \mathbb{R}^{n\times m \times d_x}$ into $\Bar{\mathbf{X}}\in \mathbb{R}^{nm\times d_x}$, where each row $\Bar{x}_{p}\,(p=1,\dots,nm)$ of the matrix $\Bar{\mathbf{X}}$ is the feature vector of a candidate item. The input of the SS-Attn module is the reshaped item matrix $\Bar{\mathbf{X}}$ and the distance matrix $\mathbf{D}$. 
To involve the page format effect (i.e., the larger the distance, the less the influence between items), we transform the distance matrix by a learnable sigmoid function~\cite{wu2021transformer}. 
The learnable sigmoid function $f$ parameterized by a scalar $v$ maps the distance $d_{p,q}$ of item pair $(p,q)$ to a positive distance-aware influence factor $\hat{d}_{p,q}$ of range $(0,1]$, $f(\cdot|v):\mathbb R^*\rightarrow \mathbb (0,1]$.
\begin{equation}
    \hat{d}_{p,q}=f(d_{p,q}|v)=\frac{1+\exp(v)}{1+\exp(v + \sigma d_{p,q})}\,,
\end{equation}
where $v\in\mathbb R$ is a learnable scalar that determines the steepness of function $f(\cdot|v)$, and $\sigma > 0$ is a hyper-parameter for normalizing the distance $d_{p,q}$ and stabilizing the training. In our experiment, $\sigma$ is set to $0.1$.  Note that $f(\cdot|v)$ is a monotonically decreasing function w.r.t. the distance and satisfies $f(0|v)=1$ and $f(+\infty|v)\rightarrow 0$, i.e., the influence of between the items gradually decreases from $1$ to $0$ as the distance grows.

Next, we use the obtained influence factor $\hat{d}_{p,q}$ to scale the pairwise mutual influence between items on the page. Multi-head attention is adopted for modeling the interactions between any pair of items on the page, while the attention weights are scaled according to the distance-aware influence factor. Suppose $B$ is the number of heads, we maintain $B$ different learnable sigmoid functions to learn different levels of the page format effect, with an individual parameter $v^{(b)}$ for the $b$-th attention head, $b=1,\ldots,B$. We form all the $\hat{d}_{p,q}$ for the $b$-th head into a matrix $\hat{\mathbf{D}}^{(b)}$, and numerically scale the preliminary self-attention weights:
\begin{equation*}
\begin{split}
    \Bar{\mathbf{O}}^{(b)}&=\operatorname{Softmax}\left(\frac{\phi((\Bar{\mathbf{X}}W_Q^{(b)})(\Bar{\mathbf{X}}W_K^{(b)})^\top)\odot \hat{\mathbf{D}}^{(b)}}{\sqrt{d_a}}\right)(\Bar{\mathbf{X}}W_V^{(b)}),
\end{split}
\end{equation*}
where $\odot$ is the element-wise product, the preliminary attention weights $(\Bar{\mathbf{X}}W_Q^{(b)})(\Bar{\mathbf{X}}W_K^{(b)})^\top$ are adjusted according to the distance-aware influence factors, and $d_a$ is the dimension of the vectors in $\Bar{\mathbf{X}}W_Q^{(b)}$
and $\Bar{\mathbf{X}}W_K^{(b)}$. Note that the non-negative monotonically increasing function $\phi$ is introduced to avoid negative attention weights, as negative preliminary attention weights can invert the distance-aware influence and violates the negative correlation between distances and influences. Here, we use $\operatorname{softplus}$ function.


Finally, we concatenate the multi-head spatial-scaled attention outputs, and apply a linear transformation to get the pairwise influence matrix $\Bar{\mathbf{O}}\in\mathbb{R}^{nm\times d_o}$, $\Bar{\mathbf{O}} = [\Bar{\mathbf{O}}^{(1)}\,\Vert\,\dots\,\Vert\,\Bar{\mathbf{O}}^{(B)}]\mathbf{W}_O$,
where $d_o$ is the attention output size. We reshape the matrix $\Bar{\mathcal{O}}\in\mathbb{R}^{nm \times d_o}$ back to $\mathbf{O}\in\mathbb{R}^{n \times m \times d_o}$, where the vector $o_{i,j} \in \mathbb{R}^{d_o}$ denotes the pairwise influence vector for the $j$-th item in the $i$-th list.

\subsubsection{Dense Network}

In addition to the HDS-Attn and SS-Attn, we employ a fully-connected network to capture the implicit feature interactions within each item. We feed each item embedding $\hat{x}_{i,j}$ into a shared MLP to obtain the dense feature $r_{i,j}=\operatorname{MLP}(\hat{x}_{i,j})$ for the latter reranking score estimation.
\vspace{-5pt}
\subsection{Reranking Score Estimation}

After the page-level interaction layer, we obtain the shared page representation $S$, the pairwise item influence vector $o_{i,j}$, and the dense feature $r_{i,j}$ for item $x_{i,j}$. Although these features incorporate page and item-level information, the commonalities and distinctions of user behaviors on different lists are remained to be solved.

As discussed before (\textbf{C3} in Section~\ref{sec:intro}), the user's behaviors may not only share some basic patterns (e.g., position bias) and underlying preferences, but also have distinctions for different lists due to themes and formats. To this end, we adopt the Multi-gated Mixture-of-Experts (MMoE~\cite{ma2018modeling}) module, where several expert sub-models are shared across all lists. Each list possesses a specific gating network to "select" a subset of experts to
use. Through the expert and gating networks, our model automatically adjusts parameterization between modeling shared information and list-specific information, so as to exploit the common behavior patterns while paying attention to the list-specific patterns.

As shown in Figure~\ref{fig:framework}, there are $E$ parallel expert networks $\{e_k(\cdot)\}_{k=1}^{E}$, which are all MLPs with ReLU activations, to capture different aspects of behavior patterns. For each list $i$, we maintain a separate fully-connected gating network $g_i(\cdot)$ to learn a linear combination of the expert outputs $\gamma_{i,j}\in\mathbb{R}^{E}$, with $\gamma_{i,j,k}$ being the $k$-th element of $\gamma_{i,j}$.
To preserve the list-specific information, we feed the combined feature vector $\hat{z}_{i,j}$ into a list-specific tower network $t_i(\cdot)$ to get the final score $\hat{y}_{i,j}$ for the $j$-th item in the $i$-th list. 
\begin{equation}
\begin{split}
    \gamma_{i,j} &=\operatorname{Softmax}(g_i([S \,\Vert\, r_{i,j} \,\Vert\, o_{i,j}]))\,,\\
    \hat{z}_{i,j}&=\sum\nolimits_{k=1}^{E} \gamma_{i,j,k} \times e_k([S \,\Vert\, r_{i,j} \,\Vert\, o_{i,j}])\,,\\
    \hat{y}_{i,j} &= t_i(\hat{z}_{i,j})\,. \\
\end{split}
\end{equation}

We sort items in each list by the scores $\hat{y}_{i,j}$ to get the final rerankings.
Given the click label matrix $\mathbf{Y}$ of size $n\times m$ where $y_{i,j}$ denotes the click signal for the $j$-th item in the $i$-th list, we optimize the model via binary cross-entropy loss on the training page set $\mathcal P$:
\begin{equation}
    \mathcal{L}=\sum_{\mathcal{P}}\sum_{i=1}^n\sum_{j=1}^m y_{i,j}\log \hat{y}_{i,j} + (1-y_{i,j})\log(1-\hat{y}_{i,j})\,.
\end{equation}
\textbf{Computational Complexity Analysis.} The complexity of PAR is $\mathcal O(n^2m^2)$, where $n$ is the number of lists and $m$ is the length of the lists. Its complexity is comparable to most existing models of  $\mathcal O(nm^2)$ (e.g., PRM~\cite{pei2019personalized}, DHANR~\cite{ding2019whole}), as the number of lists on the page $n$ is usually less than 5 due to the limit of the screen size. 

\vspace{-5pt}
\section{Experiment}
\label{sec:exp}
\subsection{Experiment Settings}
\subsubsection{Datasets}\label{sec:dataset} 
Our experiments are conducted on a public dataset, Cloud Theme Click Dataset\footnote{https://tianchi.aliyun.com/dataset/dataDetail?dataId=9716}, and a proprietary dataset, AppStore.
\begin{itemize}[leftmargin=10pt]
    \item \textbf{Cloud Theme Click Dataset}~\cite{du2019sequential} (\textbf{CTC} for short) records the click data of Cloud Theme in Taobao app. The dataset includes 1,423,835 click records from 355 different themes during a 6-day promotion season, users'  purchase history before the promotion, and the embedding of 720,210 users and 1,361,672 items. 
    \item \textbf{AppStore} is collected from a mainstream commercial app store, from October 16, 2021 to November 1, 2021. The dataset contains 47,003,121 pages, 28,632,998 users and 1365 apps. Each app has 32 features (e.g., app developer, app category). Each page is in the form of an F-shape with one vertical list inserted by four horizontal lists, and each user has a history list of behaviors collected in real time.
\end{itemize}
\vspace{-5pt}
\subsubsection{Page and click generation for public dataset}\label{page_click} As there is \textbf{no publicly available} dataset with page data, we use the click logs of the public CTC dataset to construct the pages. For each user, we first construct lists from her/his positively interacted themes with at least one click. Next, each page is formed by four lists from different themes that are horizontally stacked from top to bottom. We randomly sample lists from all the themes if less than four lists have been clicked. Four DNN models are trained separately as the initial rankers to generate the initial rankings of length 10. 


An oracle click model (e.g.,~\cite{lin2021graph,dai2021adversarial}) is then adopted to simulate necessary click data on the obtained new pages. We mainly follow the click simulation in Seq2Slate \cite{bello2018seq2slate} to decompose the click probability of an item into the product of relevance, position decaying, and dissimilarity probability. We use the original click label of an item as the relevance probability, which is equal to $1$ if clicked, and $0$ otherwise. If the item is placed at $i$-th position in the $j$-th list, the position decaying probability is $1/(i^{\eta_1}j^{\eta_2})$, where $\eta_1$ and $\eta_2$ are the horizontal and vertical decay parameters. Here we set $\eta_1$ and $\eta_2$ to 0.4 and 0.5. When observing an item, the user may tend to click the item dissimilar/diverse to its surrounding items~\cite{diverOrNot, clkToDiver}. The cosine similarity between item embeddings is used to compute the dissimilarity probability to introduce high-order interaction between items. After the generation of clicks, all baselines and our model PAR are trained based on the synthetic click data.
\vspace{-5pt}
\subsubsection{Baselines} 
Since currently there is only one work, \textbf{DHANR}, that focuses on page-level reranking, we modify some models in related fields that utilize page-level information as baselines, e.g.,  \textbf{HMoE} in multi-scenario ranking, \textbf{TRNN} in whole-page optimization. We also design \textbf{GlobAtt} to directly model the influences between items on a page. The relationship and differentiation between page-level reranking and the related fields are discussed in Section~\ref{sect:related}.
\begin{itemize}[leftmargin=10pt]
    \item \textbf{DHANR}~\cite{hao2021re} applies a hierarchical attention network to aggregate the item features into a unified static page representation, which is shared for the list-level reranking in each list.
    \item \textbf{HMoE}~\cite{li2020improving} adopts MMoE to implicitly identify distinctions and commonalities among lists from different scenarios.
    \item \textbf{TRNN}~\cite{lo2021page} focuses on whole-page optimization and ranks the widgets (fixed ordered lists) by a two-stage RNN. Necessary modifications to TRNN are made so that it can employ the page information to help rerank the items in different lists.
    \item \textbf{GlobAtt} fuses multiple lists into one mixed list and adopts a global multi-head self-attention structure to model the mutual influence between any pairs of items on the page. List and position context are fed into the network as additional features.
\end{itemize}

We also deploy list-level reranking for multiple lists on the same page. Multiple models are trained separately on different lists.
\begin{itemize}[leftmargin=10pt]
    \item \textbf{miDNN}~\cite{zhuang2018globally} uses global feature extension to extract mutual influence between items in each ranking list.
    \item \textbf{GSF}~\cite{ai2019learning} employs deep neural networks (DNN) to learn multivariate scoring functions by enumerating feasible item permutations.
    \item \textbf{DLCM}~\cite{ai2018learning} applies gated recurrent units (GRU) to encode top items in the ranking lists into feature representations.
    \item \textbf{PRM}~\cite{pei2019personalized} models the mutual influence between any pairs of items and users' preferences by self-attention.
    \item \textbf{SetRank}~\cite{pang2020setrank} learns permutation-equivariant representations for the inputted documents via the self-attention structure.
\end{itemize}

\begin{table*}[]
    \caption{Overall performance on the AppStore and CTC datasets.}
    \label{tab:overall}
    \vspace{-10pt}
    \centering
\setlength{\tabcolsep}{2pt}
\begin{adjustbox}{max width=0.965\linewidth}
    \begin{tabular}{c|ccccccccc|cccccccc}
    \toprule
    \multirow{2}{*}{\textbf{Models}} & \multicolumn{9}{c}{\textbf{AppStore dataset}}  & \multicolumn{8}{c}{\textbf{CTC dataset}}  \\
    \cmidrule{2-18}
    & $\mathbf{Utility}$            & $\mathbf{sCTR}$             & $\mathbf{sCTR_v}$          & $\mathbf{sCTR_{h1}}$         & $\mathbf{sCTR_{h2}}$         & $\mathbf{sCTR_{h3}}$         & $\mathbf{sCTR_{h4}}$         & $\mathbf{nDCG}$            & $\mathbf{MAP}$             & $\mathbf{Utility}$            & $\mathbf{sCTR}$             & $\mathbf{sCTR_{h1}}$         & $\mathbf{sCTR_{h2}}$         & $\mathbf{sCTR_{h2}}$         & $\mathbf{sCTR_{h3}}$         & $\mathbf{nDCG}$             & $\mathbf{MAP}$              \\
    \midrule
    INIT                             & 1.0459           & 1.0571           & 0.2896           & 0.6953           & 0.0391          & 0.0122          & 0.0098          & 0.6260          & 0.5087          & 1.2706           & 1.2545           & 0.5030           & 0.3747           & 0.3161          & 0.0607          & 0.4953           & 0.3241           \\
    miDNN                            & 1.1704           & 1.1993           & 0.3428           & 0.7588           & 0.0396          & \textbf{0.0186} & 0.0107          & 0.6375          & 0.5204          & 1.3530           & 1.3305           & 0.5302           & 0.3884           & 0.3439          & 0.0680          & 0.5090           & 0.3415           \\
    GSF                              & 1.2183           & 1.1866           & 0.3970           & 0.7505           & 0.0420          & 0.0146          & 0.0142          & 0.6344          & 0.5171          & 1.2791           & 1.3252           & 0.5308           & 0.3899           & 0.3386          & 0.0659          & 0.5238           & 0.3618           \\
    DLCM                             & 1.3218           & 1.2593           & 0.4521           & 0.8052           & 0.0361          & 0.0176          & 0.0107          & 0.6234          & 0.5020          & 1.3494           & 1.3603           & 0.5371           & 0.4064           & 0.3504          & 0.0664          & 0.5303           & 0.3692           \\
    PRM                              & 1.3438           & 1.2748           & 0.4497           & 0.8242           & 0.0430          & 0.0151          & 0.0117          & \textbf{0.6403}          & \textbf{0.5238} & 1.3722           & 1.3710           & 0.5474           & 0.4014           & 0.3539          & 0.0684          & 0.5212           & 0.3578           \\
    SetRank                          & 1.3535           & 1.2938           & 0.4863           & 0.8013           & 0.0366          & 0.0166          & 0.0127          & 0.6388   & 0.5223          & 1.3643           & 1.3635           & 0.5432           & 0.4026           & 0.3507          & 0.0671          & 0.5300           & 0.3689           \\ \midrule
    GlobAtt	& 1.3896 & 1.3611 & 0.5335 &	0.7586 &	0.0415 &	0.0169 &	0.0107 & 0.6245 &	0.5033 &	1.3678 &	1.3638 &	0.5449 & 	0.4039 &	0.3474 & 	0.0676 &	0.5185  & 0.3538\\
    TRNN                             & 1.3970           & 1.3641           & 0.5029           & 0.7899           & 0.0431          & 0.0172          & 0.0111          & 0.6351          & 0.5172          & 1.3750           & 1.3602           & 0.5533           & 0.3965           & 0.3430          & 0.0674          & 0.5220           & 0.3587           \\
    HMoE                             & 1.3706           & 1.3744           & 0.5394           & 0.7661           & 0.0414          & 0.0167          & 0.0108          & 0.6373          & 0.5203          & 1.3615           & 1.3608           & 0.5452           & 0.3968           & 0.3492          & \textbf{0.0696} & 0.5236           & 0.3609           \\
    DNAHR                            & 1.4116           & 1.3872           & 0.5312           & 0.8213           & 0.0405          & 0.0107          & 0.0078          & 0.6312          & 0.5121          & 1.3303           & 1.3512           & 0.5382           & 0.3992           & 0.3450          & 0.0687          & 0.5233           & 0.3597           \\
    \textbf{PAR}                     & \textbf{1.5024*} & \textbf{1.4457*} & \textbf{0.5801*} & \textbf{0.8413*} & \textbf{0.0503} & 0.0161          & \textbf{0.0146} & 0.6333          & 0.5151          & \textbf{1.4446*} &	\textbf{1.4137*} &	\textbf{0.5693*} &	\textbf{0.4174*} &	\textbf{0.3594} &	0.0677	 & \textbf{0.5548*} &	\textbf{0.4005*}\\
    \bottomrule
    \end{tabular}
\end{adjustbox}
\footnotesize \flushleft\hspace{0.3cm} $*$ denotes statistically significant improvement (measured by t-test with $p$-value $<$ 0.05) over the best baseline.
\end{table*}

\subsubsection{Evaluation metrics} All the reranking models are evaluated in terms of relevance-based and utility-based metrics. For relevance-based metrics, we adopt widely-used \textit{MAP} and \textit{nDCG}~\cite{ndcg} following previous work \cite{lo2021page,pei2019personalized, ai2018learning}. Despite that there are multiple lists on the page, we calculate \textit{nDCG} and \textit{MAP} for each list and report their averaged \textit{nDCG} and \textit{MAP}.

As for utility-based metrics, we employ the average number of clicks on the page \textit{Utility}, and the \textit{sum} of the click probabilities for all items on a page \textit{sCTR}, following~\cite{xi2021context}. In addition, we also compute the sum of click probabilities on each list, such as the sum of click probabilities on vertical lists $sCTR_v$, and the sum of click probabilities on the first horizontal lists $sCTR_{h1}$. On the public CTC dataset, the click probabilities and clicks of the reranked lists are generated by the same oracle click model used in \ref{page_click}. The AppStore dataset records the real user clicks on the F-shape page. Its click probabilities and clicks on the reranked lists are given by a click model for the F-shape page, FSCM~\cite{fu2022f}.


\subsubsection{Reproducibility}
The implementation of our model is available\footnote{The TensorFlow implementation is available at: \url{https://github.com/YunjiaXi/Page-level-Attentional-Reranking}. The MindSpore implementation is available at: \url{https://gitee.com/mindspore/models/tree/master/research/recommend/PAR}}.
We adopt Adam as the optimizer. The learning rate is 
$2\times10^{-4}$, and the parameter of $l_2$ regularization is $2\times 10^{-4}$. The batch size and the embedding size of the categorical feature are set to 128 and 16. The number of experts and the architecture of experts and towers in MMoE are 12, [200, 80], and [80], respectively. To ensure a fair comparison, we also fine-tune all baseline models to achieve their best performance. 


\subsection{Overall Performance}
The overall performance of our proposed PAR and baselines on the AppStore and CTC datasets are reported in Table \ref{tab:overall}, from which we have the following observations.

\textit{ Firstly, our model PAR performs significantly better than all the baselines on both datasets.} Methods with page-wise information such as HMoE, TRNN, DHANR, and GlobAtt generally work better than list-level methods, validating the benefit of utilizing page-level information. PAR surpasses all the baselines on the two datasets. As presented in Table~\ref{tab:overall}, PAR improves over the best baseline on CTC dataset with respect to \textit{Utility}, \textit{sCTR}, \textit{nDCG}, and \textit{MAP} by 5.276\%, 3.115\%, 4.623\%, and 8.485\%, respectively. On AppStore dataset, PAR also achieves 6.432\% and 4.217\% improvement over the best baseline in terms of \textit{Utility} and \textit{sCTR}.
This demonstrates the necessity of modeling the multifaceted dynamic interactions and the page format in page-level reranking. 

\textit{Secondly, different page formats result in different click distributions. }As illustrated in Table~\ref{tab:overall}, the top clicks are predominantly located on the vertical and the first horizontal lists on the F-shape pages of AppStore dataset. On the all-row pages of CTC dataset, clicks are concentrated on the first three horizontal lists and the probability of clicking decreases with the positions of the horizontal list. PAR shows greater improvement in these major lists by exploiting more useful information. Page-level baselines outperform list-level models on AppStore dataset, but sometimes this is not the case for the CTC dataset, which may also be due to the different page formats. The inter-list interaction for pages of multiple horizontal lists may be less than that for F-shape pages.

\textit{Lastly, the performance of the models on the AppStore and CTC datasets diverges in terms of the relevance-based metrics, \textit{nDCG} and \textit{MAP}.} On the CTC dataset, PAR achieves the best nDCG and MAP, but on the AppStore dataset, the best relevance-based metrics are achieved by the list-level reranking method, PRM. 
This may be attributed to the fact that CTC dataset has groundtrue relevance labels, whereas the AppStore does not.
The click labels in AppStore dataset are directly used for computing the relevance-based metrics, following~\cite{lo2021page,pei2019personalized, ai2018learning}. Yet the clicks could be biased, and there exist some relevant items that have not been clicked. The list-level approaches tend to place the past-clicked items first, and therefore obtain higher nDCG and MAP. In comparison, the page-level approaches combine information from multiple lists to find relevant items that might not have been clicked to optimize the total utility and often do not necessarily place clicked items at top positions. Such an observation shows that using past clicks for evaluation may not be able to reflect the true performance for page-level reranking.

\begin{table*}[]
    \caption{Ablation on AppStore and CTC datasets.}
    \label{tab:ablation}
    \vspace{-10pt}
    \centering
\setlength{\tabcolsep}{2.5pt}
\begin{adjustbox}{max width=0.96\linewidth}
\begin{tabular}{c|ccccccccc|cccccccc}
\toprule
\multirow{2}{*}{\textbf{Variants}} & \multicolumn{9}{c|}{\textbf{AppStore dataset}} & \multicolumn{8}{c}{\textbf{CTC dataset}} \\
\cmidrule{2-18}
& $\mathbf{Utility}$            & $\mathbf{sCTR}$             & $\mathbf{sCTR_v}$          & $\mathbf{sCTR_{h1}}$         & $\mathbf{sCTR_{h2}}$         & $\mathbf{sCTR_{h3}}$         & $\mathbf{sCTR_{h4}}$         & $\mathbf{nDCG}$            & $\mathbf{MAP}$             & $\mathbf{Utility}$            & $\mathbf{sCTR}$             & $\mathbf{sCTR_{h1}}$         & $\mathbf{sCTR_{h2}}$         & $\mathbf{sCTR_{h2}}$         & $\mathbf{sCTR_{h3}}$         & $\mathbf{nDCG}$             & $\mathbf{MAP}$              \\
\midrule
PAR-DSA & 1.4482 & 1.4236 & 0.5515 & 0.8009 & 0.0432 & 0.0171 & 0.0110 & 0.6292 & 0.5097 & 1.3878 & 1.3882 & 0.5493 & 0.4133 & 0.3584 & 0.0673 & 0.5379 & 0.3789 \\
PAR-HDSA & 1.4129 & 1.4193 & 0.5566 & 0.7917 & 0.0431 & \textbf{0.0172} & 0.0108 & 0.6337 & 0.5154 & 1.3629 & 1.3757 & 0.5593 & 0.4113 & 0.3406 & 0.0643 & 0.5334 & 0.3736 \\
PAR-scale    & 1.4136          & 1.4356          & 0.5547          & 0.8091          & 0.0436          & \textbf{0.0172} & 0.0109          & 0.6305          & 0.5116          & 1.3895          & 1.3933          & 0.5589          & 0.4095          & 0.3556          & \textbf{0.0692} & 0.5399          & 0.3816          \\
PAR-SSA      & 1.4595          & 1.4338          & 0.5570          & 0.8083          & 0.0414          & 0.0165          & 0.0105          & 0.6328      & 0.5143          & 1.3885          & 1.3914          & 0.556           & 0.4092          & 0.3586          & 0.0676          & 0.5393          & 0.3809          \\
PAR-DN     & 1.4743          & 1.4343          & 0.5555          & 0.8076          & 0.0431          & 0.0171          & 0.0110          & 0.6327          & 0.5144          & 1.4219          & 1.4026          & 0.5657          & 0.4151          & 0.3552          & 0.0666          & 0.5423          & 0.3848          \\
PAR-MMoE & 1.3950 & 1.4203 & 0.5501 & 0.7997 & 0.0427 & 0.0169 & 0.0108 & \textbf{0.6349} & \textbf{0.5171} & 1.3716 & 1.3751 & 0.5531 & 0.4056 & 0.3484 & 0.068 & 0.5279 & 0.3658\\
\textbf{PAR} & \textbf{1.5024} & \textbf{1.4457} & \textbf{0.5801} & \textbf{0.8413} & \textbf{0.0503} & 0.0161          & \textbf{0.0146} & 0.6333          & 0.5151          & \textbf{1.4446} & \textbf{1.4137} & \textbf{0.5693} & \textbf{0.4174} & \textbf{0.3594} & 0.0677          & \textbf{0.5548*} & \textbf{0.4005*}\\
\bottomrule
\end{tabular}
\end{adjustbox}
\footnotesize \flushleft\hspace{0.3cm} $*$ denotes statistically significant improvement (measured by t-test with $p$-value $<$ 0.05) over the best variant.
\end{table*}

\vspace{-5pt}
\subsection{In-depth Analysis}

\subsubsection{Ablation study}
To investigate the effectiveness of each component in PAR, we design several variants of PAR and conduct a series of experiments on AppStore and CTC datasets.
\begin{itemize}[leftmargin=10pt]
    \item \textbf{PAR-DSA} replaces the dual-side attention in HDS-Attn with a self-attention, thus removing the user history from the module.
    \item \textbf{PAR-HDSA} removes the HDS-Attn module.
    \item \textbf{PAR-scale} replaces the SS-Attn module with  self-attention.
    \item \textbf{PAR-SSA} removes the SS-Attn module.
    \item \textbf{PAR-DN} removes the dense network.
    \item \textbf{PAR-MMoE} replaces the MMoE module with a single MLP.
\end{itemize}
The comparison of the above variants and the original PAR on AppStore and CTC datasets are presented in Table~\ref{tab:ablation}. Compared to the original PAR, the performances of the variants all decline to some extent, indicating the effectiveness of each module. Among all the variants, PAR-HDSA generally suffered the greatest drop in utility, which suggests that incorporating intra- and inter-list interaction can enhance the performance of page-level reranking. The decline of PAR-DSA indicates the importance of personalized preferences in user behaviors for reranking. Removing the MMoE and SSA modules also introduces a large decrease, illustrating the impact of modeling the differences and commonalities between lists and the page format. There is no significant gap between the results of PAR-SSA and PAR-scale, revealing that it is the distance-aware influence factor that contributes primarily to the SSA module. The basic self-attention is insufficient to model different page formats.

\subsubsection{Case Study}

 \begin{figure}[t]
    \centering
    \vspace{0pt}
    \includegraphics[width=\linewidth]{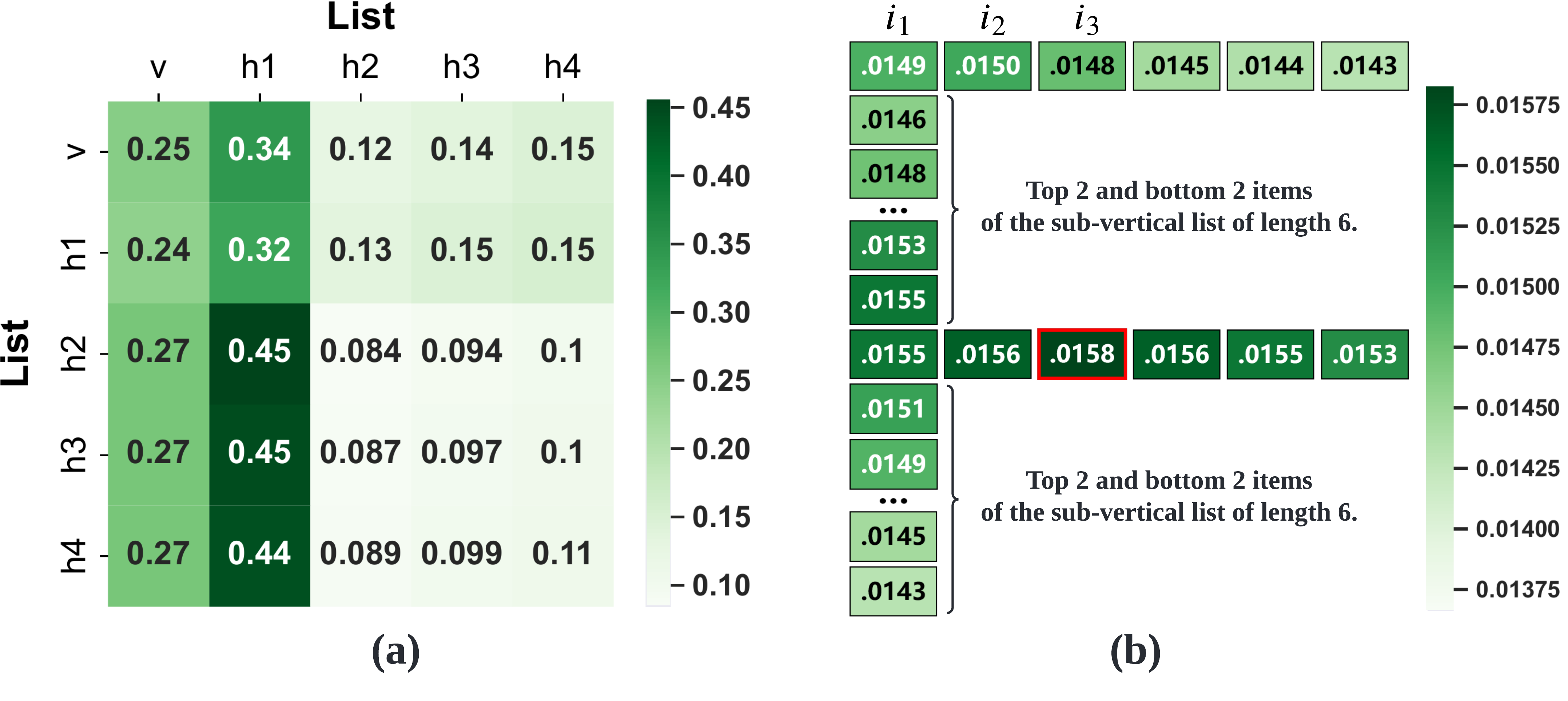}
    \vspace{-20pt}
    \caption{
    (a) The attention weights (normalized along the horizontal axis) of the list-level self-attention in HDS-Attn. 
    (b) The pairwise attention weights between the target item (highlighted by a red box) and other items in SS-Attn.}
    \label{fig:case_study}
    \vspace{-13pt}
\end{figure}

To explore the mutual influence between lists and the spatial relationship between items, we select a page from the AppStore and visualize the attention weights from list-level self-attention in HDS-Attn and spatial-scaled attention (SS-Attn). Figure~\ref{fig:case_study}(a) presents the heatmap of the vertical list $v$ and the four horizontal lists $h1$, $h2$, $h3$, and $h4$, with each row being the attention weights of a list. In Figure~\ref{fig:case_study}(a), all lists have a high attention weight for $h1$, probably because $h1$ has the most clicks and can provide more information. List $v$, neighboring all the other lists, receives the next highest number of clicks, and thus the other lists also have a relatively high attention weight for $v$. Conversely, $h2$-$h4$, located further away with lower CTR, have smaller attention weights. 

As for spatial-scaled attention, we select a target item at the third position of the second horizontal list and then visualize its attention weights on all the items on the F-shape page in Figure~\ref{fig:case_study}(b), from which we observe a general pattern of spatial decay. The target item has the greatest attention weight on itself, and the weights on its surrounding items roughly follow the pattern that the further away the item is, the less weight it gets. Furthermore, we find that there are outliers that violate the spatial decay pattern. In the first horizontal list, the first item $i_1$ and second item $i_2$ have higher attention weights (0.0149 and 0.0150) than the third item $i_3$ (0.0148), though $i_3$ is closer to the target item. The possible reason is that $i_1$ and $i_2$ share the same category with the target item (they are all short-video apps), while $i_3$ is a search engine app. The similar category promotes the pairwise mutual influence between items. As such, we conclude that the SS-Attn can automatically learn the combination of the spatial effect and the pairwise influence.


\vspace{-5pt}
\section{Related Works}\label{sect:related}


\subsection{Reranking}


Most existing reranking methods are list-level and rerank separately for each individual list \cite{feng2021grn,xi2021context,ai2018learning,bello2018seq2slate,feng2021revisit,pang2020setrank,pei2019personalized}. Various network structures have been applied for modeling the mutual influences within the list. For example, miDNN~\cite{zhuang2018globally} uses DNN with global feature extension to capture mutual influences between items. A group-wise scoring function (GSF)~\cite{ai2019learning} is learned by enumerating all the feasible item permutations of the list. DLCM~\cite{ai2018learning} employs the gated recurrent unit (GRU) to encode the whole ranking list into the item representations. PRM~\cite{pei2019personalized} and SetRank~\cite{pang2020setrank} adopt the self-attention mechanism to model the influence between any pair of items in the list. Yet the performance of list-level reranking algorithms is usually suboptimal when the recommendation page presented to the user is in a multi-list style.

\citet{hao2021re} find the information from other lists on the page can improve the performance of reranking, and propose a deep and hierarchical attention network reranking (DHANR) model. They aggregate the page-wise context into a static page representation vector, and apply an identical list-level reranking algorithm for all the lists with the page representation as shared side information. However, DHANR fails to capture the dynamic page-level interaction between items, and is insensitive to the different page formats or the commonalities and distinctions between lists. Our proposed PAR fully exploits the page-wise context and captures multifaceted fine-grained item influences across lists.

\vspace{-7pt}
\subsection{Multi-scenario Learning to Rank}
Multi-scenario learning to rank aims to improve the overall performance in different scenarios, which can be generally classified into two categories: (1) multi-task learning (MTL) \cite{chen2020scenario,li2020improving,gu2021self,sheng2021one,zhao2019recommending}, and (2) multi-agent reinforcement learning (MARL) \cite{feng2018learning,he2020learning}.
MTL-based methods formulate the ranking problems from different scenarios as different tasks, and devise a single model to solve the multiple tasks simultaneously. For instance, HMoE~\cite{li2020improving} utilizes Multi-task Mixture-of-Experts~\cite{ma2018modeling,zhao2019recommending} to identify the commonalities and distinctions between scenarios implicitly. As for MARL-based methods~\cite{feng2018learning,he2020learning}, each scenario has a local ranking agent, and the agents are trained collaboratively to improve the overall performance. However, the agents are updated in an online setting with instant feedback, which is different from ours.

Page-level reranking intends to simultaneously rerank lists on the same page, which emphasizes on user behaviors when examining a recommendation page as a whole. How to model the page format and contexts is key to page-level reranking, which is ignored in multi-scenario learning to rank. 
\vspace{-7pt}
\subsection{Whole-page Optimization}
Whole-page optimization focuses on improving the display of the recommendation page, which generally falls into two categories. 

The first category~\cite{wang2016beyond,wang2018optimizing,hill2017efficient,wang2017efficient,xiao2022tile,zhao2018deep,oosterhuis2018ranking} aims to find the optimal presentation style for each item on the page. Presentation style includes positions, image sizes, text fonts, etc. The problem is then formulated as a combinatorial optimization problem of determining positions and other presentation styles for each item, where graph matching~\cite{wang2016beyond,wang2018optimizing}, bandit~\cite{hill2017efficient,wang2017efficient}, or reinforcement learning (RL)~\cite{wang2018optimizing,xiao2022tile,zhao2018deep,oosterhuis2018ranking} algorithms are proposed. This type of method, however, is designed for mapping one single initial list into a 2D geometrical layout, which is different from our work --- the input of PAR is multiple initial lists of different themes.

The second category~\cite{gomez2015netflix,bendada2020carousel,ding2019whole,lo2021page} is designed to select and rank the widgets (list of items) of the page. 
Given a set of widgets where the order of items for each widget is fixed, these methods try to select personalized themes of widgets to meet users' needs. 
These models, though involving page-level information, are not the focus of our work. In this work, we jointly optimize the arrangement of items for multiple lists by considering the page-wise context.

\vspace{-7pt}
\section{Conclusion}
In this work, we study the problem of page-level reranking, which requires a unified model to rerank multiple lists simultaneously on the same recommendation page. We conduct a data-driven study based on a real-world multi-list dataset, and propose a novel Page-level Attentional Reranking (PAR) model. We design a hierarchical dual-side attention module and a spatial-scaled attention network to learn the fine-grained spatial-aware item interactions across lists. Besides, we adopt the multi-gated mixture-of-experts module to capture the commonalities and distinctions of user behaviors among different lists.
Extensive experiments show that PAR significantly outperforms the state-of-the-art baselines.
\section*{Acknowledgement}

The SJTU team is supported by Shanghai Municipal Science and Technology Major Project (2021SHZDZX0102) and National Natural Science Foundation of China (62177033). The work is also sponsored by Huawei Innovation Research Program.
We thank MindSpore~\cite{mindspore} for the partial support of this work.

\bibliographystyle{ACM-Reference-Format}
\bibliography{acmart}

\end{document}